\documentclass[traditabstract]{aa}
\usepackage{graphicx}
\usepackage{natbib}
\usepackage{txfonts}
\begin{document}

   \title{${\rm ^{22}Ne}$ distillation and the cooling sequence of the old metal-rich open cluster NGC~6791}


   \author{Maurizio Salaris\inst{1,2} 
   \and Simon Blouin\inst{3} 
    \and Santi Cassisi\inst{2,4} 
    \and Luigi R. Bedin\inst{5}
          }

   \institute{
   Astrophysics Research Institute, Liverpool John Moores University, 146 Brownlow Hill, Liverpool L3 5RF, UK (M.Salaris@ljmu.ac.uk)
   \and
   INAF-Osservatorio Astronomico d'Abruzzo, via M. Maggini, sn.
     64100, Teramo, Italy   
   \and
   Department of Physics and Astronomy, University of Victoria, Victoria BC V8W 2Y2, Canada
   \and 
    INFN -- Sezione di Pisa, Largo Pontecorvo 3, 56127 Pisa, Italy
   \and 
    INAF -- Osservatorio Astronomico di Padova, Vicolo dell'Osservatorio 5, I-35122 Padova, Italy
     }

   \date{Received ; accepted }

 
\abstract{
Recent Monte Carlo plasma simulations to study in crystallizing carbon-oxygen (CO) white dwarfs (WDs)
the phase separation of $^{22}{\rm Ne}$ (the most abundant metal after carbon and oxygen) have shown that, under the right conditions, a distillation process that transports $^{22}{\rm Ne}$ toward the WD centre is efficient and releases a considerable amount of gravitational energy that can  
lead to cooling delays of up to several Gyr.
Here we present the first CO WD stellar evolution models that self-consistently include the effect of neon distillation, and 
cover the full range of CO WD masses, for a progenitor metallicity twice-solar appropriate for the old open cluster NGC~6791. The old age (about 8.5~Gyr) and high metallicity of this cluster --hence the high neon content (about 3\% by mass) in the cores of its WDs-- maximize the effect of neon distillation in the models to be compared with the observed cooling sequence.
We discuss the effect of distillation on the internal chemical stratification and cooling time of the models, confirming that distillation causes cooling delays up to several Gyr, that depend in a non-monotonic way on the mass. We also show how our models produce luminosity functions (LFs) that can match the faint end of the observed WD LF in NGC~6791, for ages consistent with the range determined from a sample of cluster's   eclipsing binary stars, and the main sequence turn-off. Without the inclusion of distillation the theoretical WD cooling sequences reach too faint magnitudes compared to the observations. 
We also propose {\sl James Webb Space Telescope} observations that can independently demonstrate the efficiency of neon distillation in the interiors of NGC~6791 WDs, and help resolve the current uncertainty on the treatment of the electron conduction opacities for the hydrogen-helium envelope of the WD models.
}

\titlerunning{${\rm ^{22}Ne}$ distillation and NGC~6791 WDs}
\authorrunning{Salaris M. et al.}
   \maketitle
%

\section{Introduction}\label{intro}

White dwarf stars (WDs) with either carbon-oxygen (CO) or oxygen-neon (ONe) cores are the most common end-stage of the evolution of single stars with initial mass up to $\sim$8-9$M_{\odot}$.  
Given the current age of the universe and the shape of the stellar initial mass function, the overwhelming majority of existing WDs  
produced by single stars (and non-interacting binaries) 
are the progeny of objects  
with initial masses typically between $\sim$0.8-1.0, and $\sim$6-7$M_{\odot}$, and have a CO core.

The evolution of WDs is a cooling process \citep[see, e.g., the review by][]{saumon22}, which produces a well-defined relationship between a WD luminosity and its cooling age, that can be employed as a useful cosmic clock.
Indeed in the last 2-3 decades, theoretical models of CO-core WDs together 
with photometric, spectroscopic, and 
asteroseismic data, have been employed to constrain the 
star formation history of the local disk from its WD population 
\citep[e.g.,][]{winget, oswalt, torres16, kilic17, tononi,cuk}, the ages of several open clusters
\citep[e.g.,][]{m67richer,vonHippel,ngc6791,m67,garciaberro,ngc2158,ngc6819} and 
globular clusters
\citep[e.g.,][]{m4hansen,ngc6397,winget3,m4,ratecooling,ngc6752}, as well as  
proxies for laboratories 
to investigate open questions in theoretical physics 
\citep[see, e.g., the review by][and references therein]{isernphys}.

The accuracy of the ages derived from WD cosmochronology rests on the accuracy of WD models, hence on the correct description of all those processes that can contribute to a WD energy reservoir. Beyond crystallization and phase separation of the CO chemical mixture \citep[see, e.g.,][and references therein]{segretain94, bd21}, there has been much interest in additional processes that can 
contribute substantially to the CO WD energy budget 
--hence cooling times--  
involving the presence of $^{22}{\rm Ne}$, the most abundant metal in the core of a CO WD after $^{12}$C and $^{16}$O (all other metals are at least about one order of magnitude less abundant than $^{22}{\rm Ne}$ in the CO cores).
This neon isotope is produced during the He-burning evolution of the WD 
progenitors and its abundance across the CO core is to a very good approximation  
constant, roughly equal (in mass fraction) to the progenitors' initial total metallicity 
\citep[see, e.g.,][and references therein]{bastiiacwd}, being typically of about 1.5-2\% for an initial solar chemical composition.

Given that $^{22}{\rm Ne}$ nuclei have a larger mass-to-charge ratio than the dominant 
$^{12}$C and $^{16}$O components, this results in a downward gravitational force on $^{22}{\rm Ne}$ and a slow diffusion towards the
centre in the liquid layers, with the release of gravitational energy 
\citep[see][]{bh01, db02,gb08,camisassa16,bastiiacwd,bauer}. This extra energy 
contribution can impact appreciably the models' cooling times, depending on the progenitor initial metallicity: higher initial metallicities imply larger $^{22}{\rm Ne}$ mass fractions and larger cooling time delays due to 
diffusion.

In a very recent development, \citet{b21} have performed state-of-the-art Monte Carlo plasma simulations to study 
the phase separation of $^{22}{\rm Ne}$ in a crystallizing CO WD. Based on their results, they revived the idea of $^{22}{\rm Ne}$ distillation \citep{isern1991,segretain1996}, where $^{22}{\rm Ne}$ is efficiently transported towards the WD centre and a potentially substantial amount of gravitational energy is released. This transport mechanism is triggered by the partial exclusion of $^{22}{\rm Ne}$ from the solid phase. Under the right conditions, this can render the solid buoyant by making it less dense than the coexisting liquid. The buoyant crystals then float up (and melt), displacing heavier liquid towards the centre of the star. This macroscopic transport of $^{22}{\rm Ne}$ can be much more efficient than $^{22}{\rm Ne}$ diffusion in the liquid phase, leading to larger effects on WD evolution with cooling delays of up to several Gyr.

While neon diffusion has been included in the computation of WD cooling models \citep[see, e.g.][]{gb08,a10,camisassa16,bastiiacwd,bauer}, the recently studied 
distillation process has yet to be modelled within full evolutionary calculations of WDs.
In this paper, we present the first calculations that include CO crystallization and phase separation, plus $^{22}{\rm Ne}$ diffusion in the liquid phase and 
distillation during CO crystallization.
We have calculated WD models (and isochrones) for an initial progenitor 
metallicity (about twice solar) appropriate to the super-solar old open cluster 
NGC~6791, which has already been studied as a test bench for the effect of neon diffusion on its WD age determination  
\citep[e.g.][]{garciaberro}. 
Our calculations have then been compared with 
the cluster cooling sequence, to assess whether models that include $^{22}{\rm Ne}$ distillation  are consistent with the observations. This comparison constitutes a strong test for the models and the efficiency of neon distillation in crystallizing WDs, given the old age of the cluster (hence the presence of WDs undergoing crystallization), its very high metallicity (hence a high $^{22}{\rm Ne}$ abundance, equal to about 3\% by mass, that maximizes the effect of distillation) and the existing very tight constraints on the cluster age and distance.

In the following Sect.~\ref{cluster} we reexamine the cluster age determination from its WDs in light of the constraints set by the study of several of its eclipsing binaries and the more recent WD calculations for the effect of $^{22}{\rm Ne}$ diffusion. 
Sect.~\ref{models} presents our new WD models that include $^{22}{\rm Ne}$ distillation and highlights the impact of this process on the models' chemical stratification and cooling times. A comparison of theoretical WD isochrones and luminosity functions calculated from our new WD models with the observed NGC~6791 counterparts follows 
in Sect.~\ref{comparison}, while in Sect.~\ref{test} we propose and describe an independent observational test for the efficiency of neon distillation in this cluster's WDs (and potentially help resolve the current uncertainties on the electron conduction opacities in the regime of WD envelopes) that makes use of the {\sl James Webb Space Telescope}.
A brief summary and conclusions close the paper in Sect. \ref{discussion}. 

\section{The cooling sequence of NGC~6791}\label{obs}
\label{cluster}

NGC~6791 is one of the richest open clusters, unusually old (age around  
8~Gyr) and very metal-rich ([Fe/H]$\sim$+0.3), with a colour-magnitude-diagram (CMD) strongly resembling that of a Galactic globular cluster.

\begin{figure}
     \centering
     \includegraphics[width=\columnwidth]{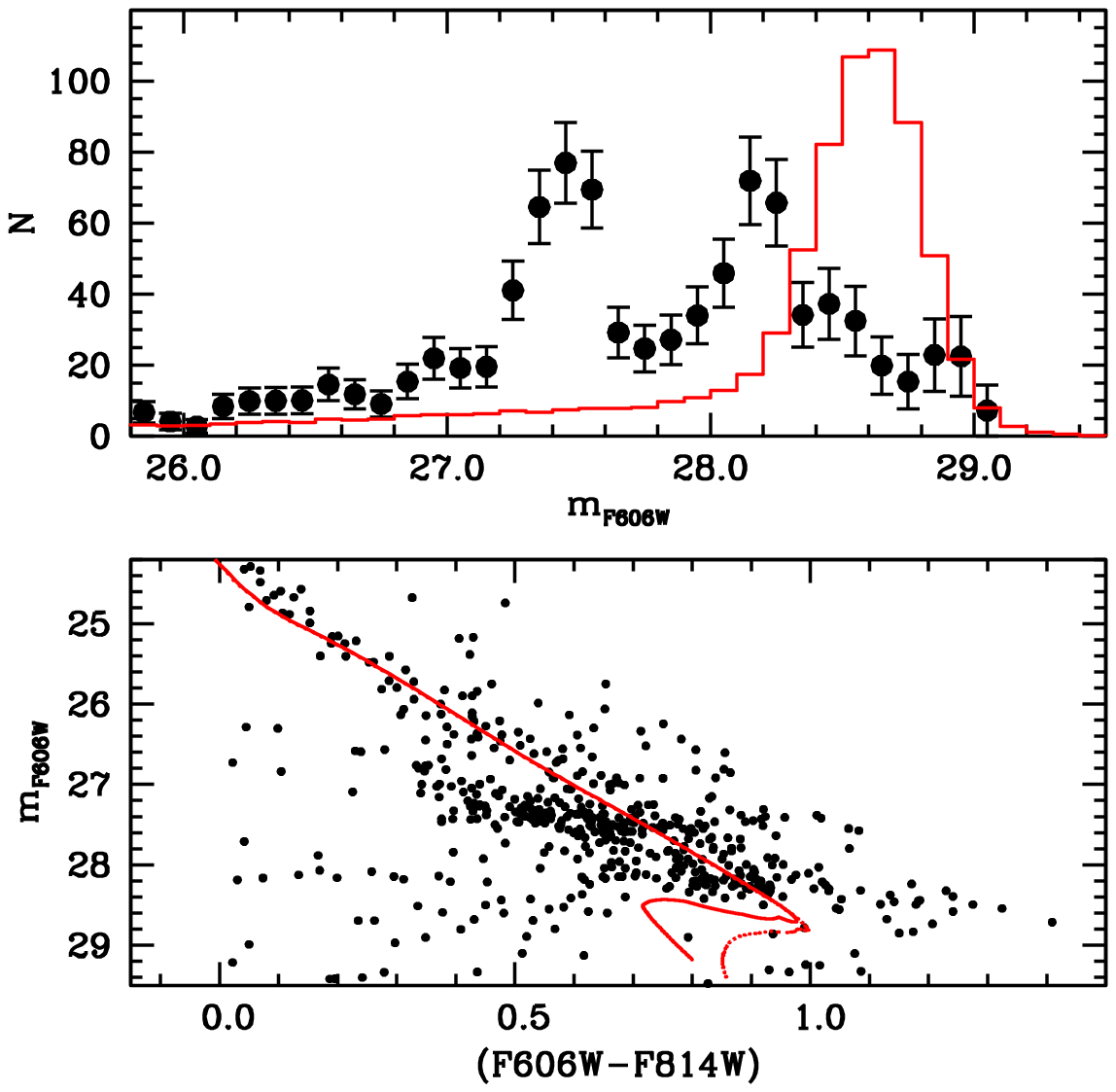}
      \caption{{\sl Upper panel:} Observed WD LF (corrected for completeness) of NGC~6791, compared with a theoretical LF calculated 
      for an age of 8.5~Gyr using a [Fe/H]=0.3 WD isochrone (with c07 opacities) from models calculated without including the effect of ${\rm ^{22}Ne}$ distillation (see text for details). {\sl Lower panel:} CMD of the observed WD cooling sequence of NGC~6791 together with [Fe/H]=0.3, 8.5~Gyr isochrones calculated without ${\rm ^{22}Ne}$ distillation and with the c07 (solid) and 
      b20 (dotted) opacities, respectively.
      }
         \label{fig:nodist}
   \end{figure}

By means of $HST$/ACS imaging, \citet{ngc6791} have reached the end of the cluster's well-populated WD cooling sequence, whose differential luminosity function (LF -- see 
Fig.~\ref{fig:nodist}) displays a peak at a magnitude
$m_{F606W}$=27.45$\pm$0.05, and fainter peak at $m_{F606W}$=28.15$\pm$0.05.
While the fainter peak is associated to the termination of the 
cooling sequence, and hence is the cluster age indicator, there is still no agreement on the origin of the brighter peak \citep[see, e.g.,][for two 
very different proposed explanations]{hansen05, ngc6791bin}.

The first analysis by \citet{ngc6791} disclosed a discrepancy between the 
cluster main sequence turn-off age and the age determined from the termination 
of the cooling sequence (the magnitude of the faint peak of the WD LF), 
the WD cluster age being lower by about 25\% (WD cluster age of about 6~Gyr against a main sequence turn-off age of about 8~Gyr), implying a too-fast 
cooling of the WD models employed in the analysis (they included CO 
crystallization and phase separation).
This discrepancy was seemingly resolved by 
\citet{garciaberro} whose WD models including also $^{22}{\rm Ne}$ diffusion had a slower cooling, and were able to recover an age of about 8~Gyr from 
the WD cooling sequence, consistent with the main sequence turn-off age.

In the intervening years, thanks mainly to the works by 
\cite{brogaard11, brogaard12, brogaard}, tight  constraints have been put to the cluster parameters 
from the analysis of the photometry and spectroscopy of a sample of its main sequence/subgiant branch eclipsing binaries, and the CMD of the cluster main sequence, red giant branch, and core He-burning phases. The resulting values for metallicity, age, reddening, and distance modulus are 
[Fe/H]=$+0.29\pm 0.03({\rm random})\pm 0.08({\rm systematic})$, 
age $t$=8.3$\pm$0.3~Gyr, 
$E(B-V)$=0.14$\pm$0.02, and $(m-M)_V$=13.51$\pm$0.06, respectively.

At the same time, the very recent model calculations by \citet{bastiiacwd} and 
\citet{bauer} have shown that the cooling delay induced by neon diffusion is 
smaller than previously calculated, leading \citet{bauer} to question whether the latest generation of WD models would still be able to lead to WD ages consistent with the cluster age determined from the main sequence.

We address this latter question by employing the recent \citet{bastiiacwd} theoretical WD cooling sequences, that include the effect of neon diffusion in addition to CO crystallization and phase separation.
Figure~\ref{fig:nodist} displays a $HST$/ACS CMD of two 8.5~Gyr WD isochrones computed with \citet{bastiiacwd} cooling tracks calculated with the 
\citet{cas07} electron conduction opacities --hereafter c07 opacities--  and the \citet{b20} opacities --hereafter b20 opacities-- respectively (see the next section for more about opacities). We used the 
\citet{cummings} WD initial-final mass relation, and progenitor lifetimes from 
\citet{bastiiacss} models\footnote{As shown in \citet{bastiiacss}, 
their evolutionary tracks and isochrones applied to study a sample of NGC~6791 eclipsing binaries and the CMD main sequence turn-off lead to an age consistent with the results mentioned in this section.}, for [Fe/H]=+0.3, corresponding to a metallicity $Z$=0.03. The WD isochrones are compared to the cluster cooling sequence \citep[from][]{ngc6791} after being shifted by $E(B-V)$=0.15 and 
$(m-M)_V$=13.50 (consistent with the constraints highlighted above), using the transformations from $A_V$ (for $R_V$=3.1) to 
$A_{F606W}$ and $A_{F814W}$ by \citet{bedin05}.

The $F606W$ magnitude of the bottom end of both isochrones is clearly fainter than the observations (the isochrone calculated with the b20 opacities is the 
faintest one, due to the shorter cooling times of the models), 
as shown also very clearly by a comparison of the theoretical LF obtained from the isochrone calculated using the c07 opacities, with the observed --completeness corrected-- counterpart. 
This theoretical LF (as well as all the other LFs that will be discussed later in 
this paper) has been calculated by drawing randomly progenitor masses according to 
a Salpeter mass function (power law with exponent equal to $-$2.35), and interpolating along the WD isochrone to determine the magnitude (and colour) of the corresponding WD. These magnitudes (and colours) were then perturbed by a Gaussian random error with $\sigma$ values determined from the photometric analysis by \citet{ngc6791}. The resulting number distribution of the magnitudes was then binned exactly as the observed LF.

Changing the age/distance modulus by $\pm$0.5Gyr/0.1~mag does not yet allow 
a match of the observed faint peak of the LF.
The bottom line is that the cooling of the WD models --for both choices of 
opacities-- seems to be too fast to match the observations, confirming 
the inference by \citet{bauer} that the cooling delay due to $^{22}{\rm Ne}$ 
diffusion might not be enough to achieve consistency between WD cluster ages 
and the age from eclipsing binaries and main sequence turn-off.

We have therefore investigated whether WD cooling tracks and isochrones including the effect of $^{22}{\rm Ne}$ distillation, can lead to a better agreement between these two independent 
age determinations.

\section{Model calculations}
\label{models}

We have computed sets of WD cooling models including the effect of $^{22}{\rm Ne}$ distillation using the same code, physics inputs, grid of masses ($M_{\rm WD}$=0.54,
0.61, 0.68, 0.77, 0.87, 1.0 and 1.1 $M_{\odot}$), and initial chemical stratification (for progenitors with [Fe/H]=0.3, corresponding to 
a metallicity $Z$=0.03, hence a uniform $^{22}{\rm Ne}$ mass fraction equal to 0.03 in the core) of the BaSTI-IAC WD calculations presented in \citet{bastiiacwd}. These latter calculations\footnote{Available at https://basti-iac.oa-abruzzo.inaf.it/} 
include $^{22}{\rm Ne}$ diffusion in the liquid phase, crystallization,  and phase separation of the CO mixture, and are our baseline models to compute the cooling delay caused by the inclusion of the distillation process. 

The models have a 
pure-H envelope that comprises a fraction $q({\rm H})=10^{-4}$ of the total mass $M_{\rm WD}$, around pure-He layers of mass fraction 
$q({\rm He})=10^{-2}$, which surround the CO core, as in the baseline calculations (the standard \lq{thick layers\rq} of H-atmosphere WD computations). As for the baseline models, we made two sets of calculations with distillation, using the c07 and b20 electron conduction opacities, respectively. We will see that quantitatively the effect of distillation on the cooling times depends on the choice of opacities. In a nutshell, these two sets of opacities treat differently the regime at the transition from moderate to strong degeneracy, which is the relevant region covered by the 
H and He envelopes of the models and it is still subject to sizable uncertainties 
\citep[see][for more in-depth discussions]{b20, cpsp, bastiiacwd}.

To include the $^{22}{\rm Ne}$ distillation we have proceeded as follows.
When, after a computational time step, the Coulomb parameter $\Gamma$ 
(expressed in terms of the Coulomb parameter for a pure carbon composition $\Gamma_{\rm C}$, see below) in one or more layers reaches the critical value $\Gamma_{\rm cr, C}$ for the crystallization of the CO mixture according to the phase diagram by 
\citet{bd21} --using Eq. 33 by \citet{bd21}, but see Eq.~\ref{gammacr} for a modification to account for the 
presence of $^{22}{\rm Ne}$-- we check whether the $^{22}{\rm Ne}$ abundance in the surrounding liquid layers is higher than the minimum value for distillation to occur, as given by \citep[from the calculations by][]{b21}:
\begin{equation}
 x_{\rm Ne} = -0.0004866  \Gamma_{\rm cr, C} + 0.09076   
\label{eqmin}
\end{equation}
where $x_{\rm Ne}$ denotes the number fraction of $^{22}{\rm Ne}$ nuclei.

If the actual abundance is above the threshold, we start 
the distillation by evolving the abundances of $^{22}{\rm Ne}$, carbon, and oxygen in these layers towards the values of the distilled composition (in mass fractions) 0.3143, 0.6857, and 0.0, respectively, which would be eventually reached when locally $\Gamma_{\rm C}$=208. The quantity $\Gamma_{\rm C}=6^{2/3} e^2/(a_e k_B T)$ denotes the value of the Coulomb 
parameter for a pure carbon composition, where $e$ is 
the elementary charge, $k_B$ the Boltzmann constant, $a_e=(4 \pi n_e/3)^{-1/3}$ the mean inter-electronic distance, $n_e$ the electron number density; when $\Gamma_{\rm C}$=208 the distillation of neon is completed and the layers where neon has accumulated do crystallize.\footnote{$\Gamma_{\rm C}$ is a dimensionless measure of $T$, 
given that $n_e$ is basically constant at constant $P$ in the degenerate core.}

While current phase diagram calculations \citep{caplan20,b21} can be used to clearly identify the starting point of the distillation process (i.e., when the solid becomes lighter than the coexisting liquid) and its stopping point (i.e., when the liquid and solid phases have the same composition), the exact trajectory followed between these two points remains uncertain. In our calculations, we make the simple assumption of a linear variation of the composition with respect to $\Gamma_{\rm C}$.

   \begin{figure}
     \centering
     \includegraphics[width=\columnwidth]{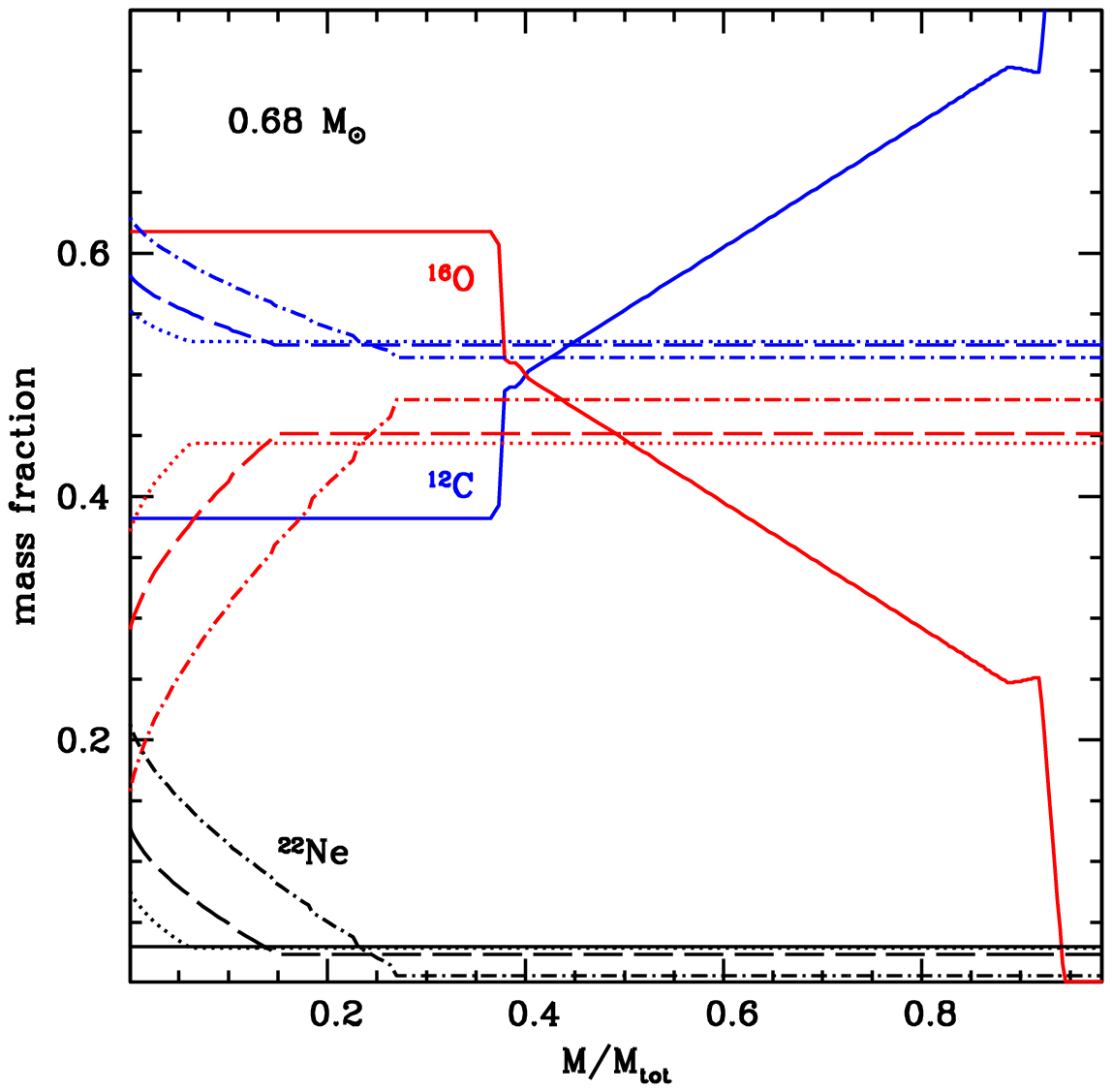}
      \caption{$^{12}{\rm C}$, $^{16}{\rm O}$, and $^{22}{\rm Ne}$ abundance profiles across 0.68$M_{\odot}$ WD models at the start of the cooling sequence (solid lines), at the end of the distillation process (dash-dotted lines), and 
      at two intermediate stages (dotted and dashed lines in order of decreasing core temperature, respectively) (see text for details).}
         \label{fig:distprog}
   \end{figure}

When the chemical abundances in the layers undergoing distillation  
are being evolved after a time step, the outer liquid layers are mixed up to the outer boundary of the core \citep[see][]{b21} and their abundances (uniform, because of the full mixing) are recalculated to ensure the conservation of the mass of the various elements -- we remark that when distillation starts at the centre we also stop the diffusion of $^{22}{\rm Ne}$ in the liquid layers.

The difference in the internal energy between the model with the new abundances and that at the previous time step is then accounted for in the energy equation.

This procedure is repeated at the following time steps and the outermost layer starting distillation moves steadily outwards with decreasing core temperature, as long as 
the $^{22}{\rm Ne}$ abundance in the liquid phase is above the required threshold.

   \begin{figure}
     \centering
     \includegraphics[width=\columnwidth]{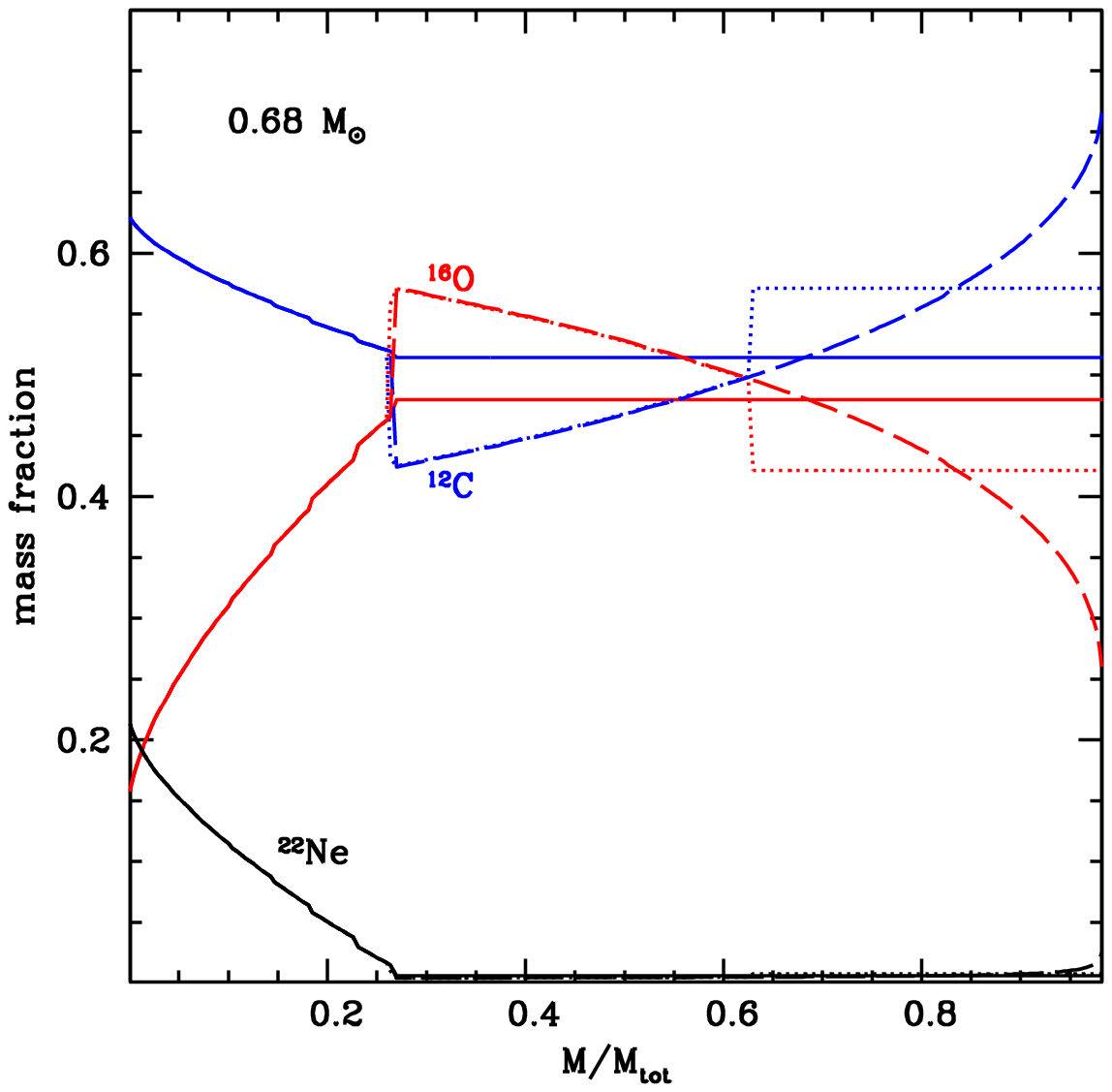}
      \caption{Chemical abundance profiles 
      across 0.68$M_{\odot}$ WD models at the 
      start of crystallization (solid lines), at the end of crystallization (dashed lines), and at an intermediate stage (dotted lines). Dotted, dashed, and solid lines 
      overlap in the inner core which has undergone neon 
      distillation. Dotted and dashed lines overlap also out to 
      $M/M_{\rm tot}\sim$0.6 (see text for details).}
         \label{fig:crystprog}
   \end{figure}

Figure~\ref{fig:distprog} displays, as an example, the progressive change of the chemical stratification during neon distillation across the core of our 
0.68$M_{\odot}$ models calculated with the c07 opacities (the progression is the same for the models calculated with b20 opacities).
After distillation starts all chemical profiles display a flat outer part, that encompasses the liquid layers fully mixed during neon distillation, and an inner part --that 
is undergoing distillation-- where 
the abundances display  
trends of increasing $^{22}{\rm Ne}$ and $^{12}{\rm C}$, and 
decreasing $^{16}{\rm O}$ when moving from the inner edge of the liquid region towards the centre.
The reason for this increase is that at any given time the value of $\Gamma_{\rm C}$ in layers progressively closer to centre is higher 
(because of increasing densities at roughly constant temperature), hence the local abundances are closer to the values corresponding to the full distillation of neon. 

When distillation progresses the neon and carbon abundances in the liquid layers decrease, and the oxygen abundance increases (as required by the conservation of the total mass of these elements), whilst the outer edge of the region undergoing distillation 
(that encloses a mass $M_{\rm dist}$) 
moves outwards and the inner abundances continue to evolve 
towards the values at full distillation.

The steadily decreasing abundance of $^{22}{\rm Ne}$ in 
the liquid phase eventually drops below the threshold for distillation to continue (given by Eq.~\ref{eqmin}) before $\Gamma_{\rm C}$=208 anywhere within $M_{\rm dist}$. When this happens, distillation is stopped, $M_{\rm dist}$ reaches its final value and at the next time step of the calculations the layers 
within $M_{\rm dist}$ are made to crystallize and release latent heat. Since distillation is stopped, no further compositional change is possible and therefore crystallization is inevitable.

We find that for all values of $M_{\rm WD}$ the final value of $M_{\rm dist}$ is equal to $\sim$ 25\% of $M_{\rm WD}$. 

The layers above $M_{\rm dist}$ also start to crystallize after distillation has stopped, depending 
on the local value of the Coulomb parameter, 
and undergo phase separation. The CO and $^{22}{\rm Ne}$ chemical profiles are now flat in these layers, with the mass fraction of $^{22}{\rm Ne}$ reduced to below 0.01, the exact 
value depending on $M_{\rm WD}$.
The CO phase separation is treated according to  \citet{bd21}
but with the following update to account for the presence of 
$^{22}{\rm Ne}$ during crystallization. In this phase, we do not reactivate neon diffusion in the liquid layers, because their abundance is so low that the effect on the cooling times is negligible 
\citep[see, e.g.,][]{bastiiacwd}.

The presence of neon affects the freezing temperature of the plasma compared to the case of a binary CO mixture. To take this effect into account, we specify the crystallization coupling parameter as (Bédard et al., submitted)
\begin{equation}
\label{gammacr}
\Gamma_\mathrm{cr,C} = \Gamma_\mathrm{cr,C}^{0} + (c_1 x_\mathrm{O} + c_2 x_\mathrm{O}^2 + c_3 x_\mathrm{O}^3) x_\mathrm{Ne},
\label{eqcry1}
\end{equation}
where $\Gamma_\mathrm{cr,C}^{0}$ is the value given by the CO phase diagram of \cite{bd21}, $x_\mathrm{O}$ and $x_\mathrm{Ne}$ are number fractions in the liquid phase, and $c_1 = 1096.69$, $c_2 = -3410.33$ and $c_3 = 2408.44$. Similarly, the separation between the liquidus and the solidus $\Delta x_{\rm O}$ is modified from Eq.~34 of \cite{bd21} as follows:
\begin{equation}
\label{deltaxo}
\Delta x_{\rm O} = a_0' x_{\rm Ne} + (a_1 + a_1' x_{\rm Ne}) x_{\rm O} + (a_2 + a_2' x_{\rm Ne}) x_{\rm O}^2 + a_3 x_{\rm O}^3 + a_4 x_{\rm O}^4 + a_5 x_{\rm O}^5,
\label{eqcry2}
\end{equation}
where $a_0' = 0.640125$, $a_1' = 2.218484$, $a_2'=-4.599227$, and the unprimed $a_i$ are given in Table~II of \cite{bd21}. Even when the fractionation of Ne is not strong enough to lead to distillation, the solid phase is still depleted in Ne. This is taken into account using the following equation (fitted to the simulation data of \citealt{b21}):
\begin{equation}
\label{deltaxne}
    \Delta x_{\rm Ne} = {\rm min} \left(0, -0.611587 x_{\rm Ne} + 0.782489 x_{\rm Ne} x_{\rm O} \right),
\label{eqcry3}
\end{equation}
where $\Delta x_{\rm Ne}$ is the difference in Ne number fraction between the solid and liquid phases, and the abundances on the right-hand side of the equation are for the liquid phase. Note that Eqs.~\ref{gammacr}--\ref{deltaxne} were fitted to simulation data in the range $0.3 \leq x_{\rm O} \leq 0.8$ and $0 \leq x_{\rm Ne} \leq 0.035$ and therefore should not be used for compositions lying outside that range\footnote{In the outermost few percent of the core mass of our models, during crystallization the 
oxygen number fraction drops below 0.30, and we 
have used the values provided by Eqs.~\ref{gammacr}--\ref{deltaxne} for 
$x_{\rm O}$=0.3. We have made a test by calculating models whereby we extrapolated these formulas to oxygen abundances below $x_{\rm O}$=0.3 when necessary. The differences in cooling times 
were negligible, due to the small amount of 
mass involved in the extrapolations.}. 

Figure~\ref{fig:crystprog} shows three snapshots of the chemical 
stratification for the same 0.68$M_{\odot}$ WD models discussed before, at the start of crystallization, at an intermediate stage through 
the process, and when the core is fully crystallized, respectively. 
Crystallization does not alter the chemical abundances within 
$M_{\rm dist}$, but the uniform CO and $^{22}{\rm Ne}$ 
chemical abundances beyond the distilled core 
undergo a progressive change with time due to 
phase separation, whose net effect is to move oxygen towards 
the interior and carbon towards the exterior of the core, 
as shown clearly by Fig.\ref{fig:crystprog}.

At the outer boundary of the distilled core region (which includes  about 25\% of the total WD mass), a discontinuity of the C and O abundances develops during the crystallization of the outer layers (due to the shape of the CO phase diagram, that causes a local increase of O and decrease of C at crystallization, see Eq.~\ref{eqcry2}), as can be seen in the figure.
The local increase of O and decrease of C cause an inversion of the 
mean molecular weight $\mu$ and an associated 
maximum density discontinuity $\Delta \rho/\rho \sim$0.001 (this value is typical for 
all WD models we have calculated). 
According to the work by \cite{blaes} this value of  
$\Delta \rho/\rho$ is however not large  
enough to break the solid lattice in the lower-$\mu$ solid layers, 
that would be therefore stable to overturning.

At intermediate stages between the beginning and the end of  crystallization, Fig.\ref{fig:crystprog} shows that the outer layers of the core (the crystallization front has reached layers 
enclosing 
about 60\% of the WD mass in the model shown) still display a uniform chemical profile, although the abundances of neon, carbon, and oxygen are different from the values at the beginning of the process, because of the 
abundance changes in the solid phase.

During the crystallization of the layers above the distilled core 
it is unclear whether $^{22}{\rm Ne}$ distillation restarts, because of an oscillatory behaviour between distillation (that 
would commence again when the $^{22}{\rm Ne}$ abundance in the liquid layers 
increases above the threshold value given by Eq.~\ref{eqmin}, 
which is affected by the value of the abundance of oxygen in the liquid layers) and no distillation (when the neon abundance decreases again below the 
threshold), due to the changes 
of neon and oxygen abundances in the liquid phase 
caused by the phase separation of the neighbouring solid layers 
(Eqs.~\ref{eqcry2} and \ref{eqcry3}). In a nutshell, phase separation tends to decrease the neon abundance in the crystallizing layers (see Eq.~\ref{eqcry3}), and as a consequence $^{22}{\rm Ne}$ increases in the surrounding liquid region, eventually 
becoming higher than the minimum value for distillation to start. However, when the layers just above the crystallization front start to experience neon distillation, the neon abundance in the liquid layers tends to decrease and distillation is shut off soon after it starts.

In some test calculations we found that if we let the distillation start again (when the conditions are met) by stopping CO crystallization and following the same procedure described before, distillation will cease almost straight 
away because the neon abundance in the liquid layers decreases  below the values given by Eq.~\ref{eqmin} after a few time steps.
The result is that these oscillations produce narrow 
mass shells (thickness on the order of 1\% of the 
total core mass) with only very small increases of 
the neon abundance caused by distillation and very minor contributions to the model's energy budget. 

In our final models, we have assumed  
that distillation does not restart after it has been completed in the central layers with mass up to $M_{\rm dist}$. The formation of the aforementioned narrow 
\lq{distillating\rq} shells --if this happens-- 
would have in any case no appreciable effects on our main conclusions because of their minor contributions to the energy budget.

   \begin{figure}
     \centering
     \includegraphics[width=\columnwidth]{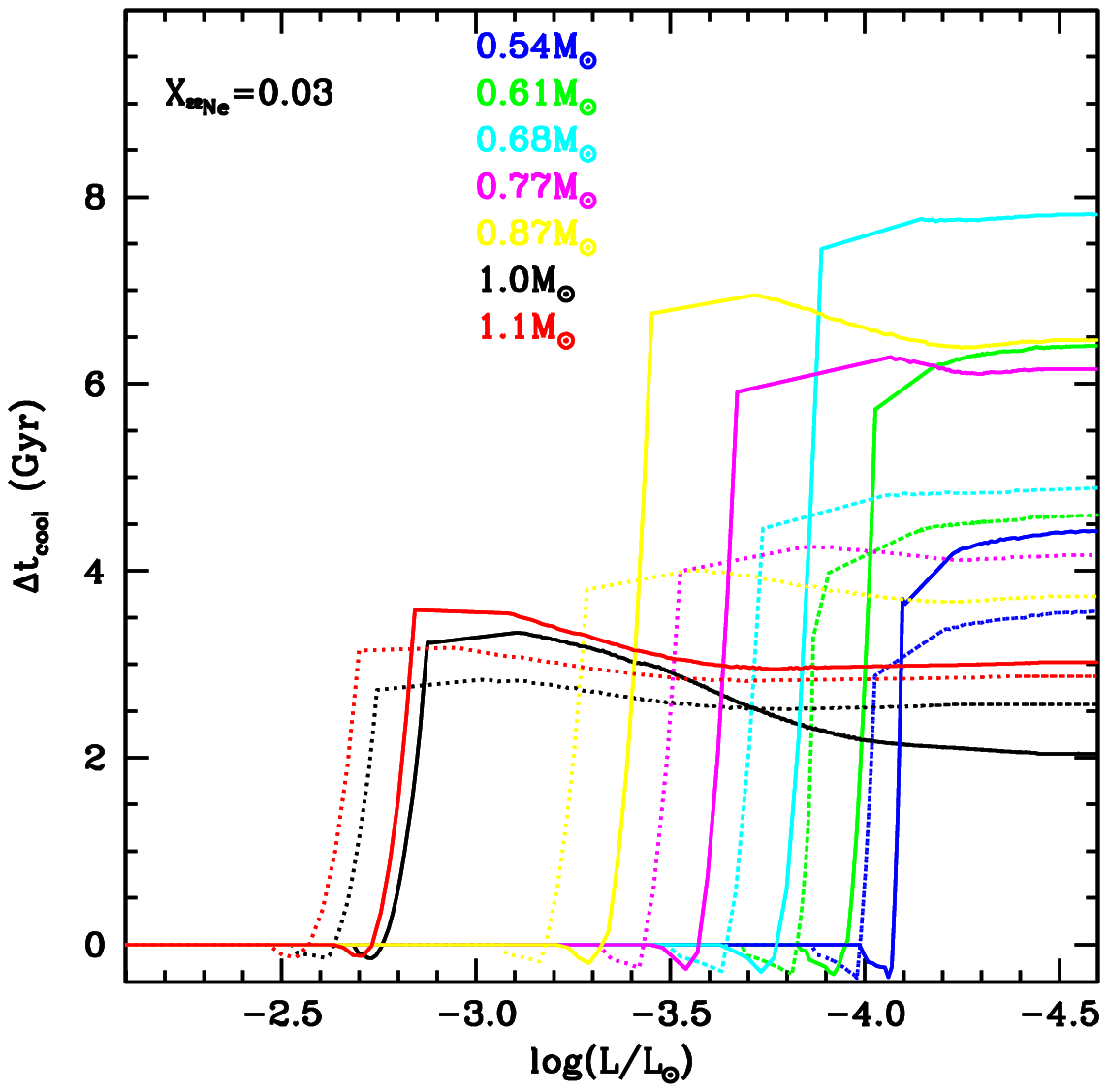}
      \caption{Difference of cooling times as a function of the luminosity between WD models --with the labelled masses-- calculated with and without ${\rm ^{22}Ne}$ distillation, using the c07 (solid lines) and b20 (dotted lines) opacities.}
         \label{fig:delayc07}
   \end{figure}


\subsection{The effect of $^{22}{\rm Ne}$ distillation on the cooling times}

Figure~\ref{fig:delayc07} displays the difference in cooling times between these new calculations and the baseline models without distillation ($\Delta {\rm t_{cool}}$), as a function of the luminosity.
We show the results for both sets of models calculated with the 
c07 opacities and b20 opacities, respectively.

When distillation first begins at the very centre of the models, 
the $\Delta {\rm t_{cool}}$ values are slightly negative because of the missing contributions from the neon diffusion in the liquid layers, and from the 
crystallization of the CO mixture in the layers that start distillation. 
This is a brief phase (it covers 
$\sim$0.1~dex in luminosity) and the $\Delta {\rm t_{cool}}$ values are never lower than $\sim -$0.3~Gyr.

With increasing time (decreasing luminosity) the effect of distillation 
overcomes the lack of energy inputs from CO crystallization and neon diffusion 
and the values of $\Delta {\rm t_{cool}}$ 
begin to increase steeply above zero. 
For all masses $\Delta {\rm t_{cool}}$ displays at this point a sharp increase with luminosity due to the energy gained during the 
distillation of $^{22}{\rm Ne}$, followed 
by a flatter profile after distillation has stopped. In this region, the exact trend of $\Delta {\rm t_{cool}}$ with luminosity depends on 
the effect of the crystallization of the CO abundance profiles  
above $M_{\rm dist}$, which are different from the baseline models, due to the chemical redistribution during distillation.

There isn't a monotonic trend of the value of $\Delta {\rm t_{cool}}$ with $M_{\rm WD}$ at the end of distillation, because the 
effect on the cooling times depends on the combination 
of how much energy is gained in the process (which increases with increasing $M_{\rm WD}$) and the luminosity at which it is released (lower masses 
start distillation at lower luminosities). For a fixed amount of 
extra energy released at a luminosity $L$, the resulting $\Delta {\rm t_{cool}}$increases with decreasing $L$.

As a result, the cooling time delay caused by the distillation of $^{22}{\rm Ne}$ first 
increases with increasing $M_{\rm WD}$, to reach a maximum for the model with $M_{\rm WD}$=0.68$M_{\odot}$, then it decreases with further increasing $M_{\rm WD}$. 
Overall the delays 
at the end of distillation are larger for the models calculated with the c07 opacities,  
that for a given value of $M_{\rm WD}$ start distillation at lower luminosities compared to the calculations 
with the b20 opacities, due to a different relationship 
between $L$ and the core temperature.
\section{Comparison with NGC~6791 cooling sequence}
\label{comparison}

\begin{figure}
     \centering
     \includegraphics[width=\columnwidth]{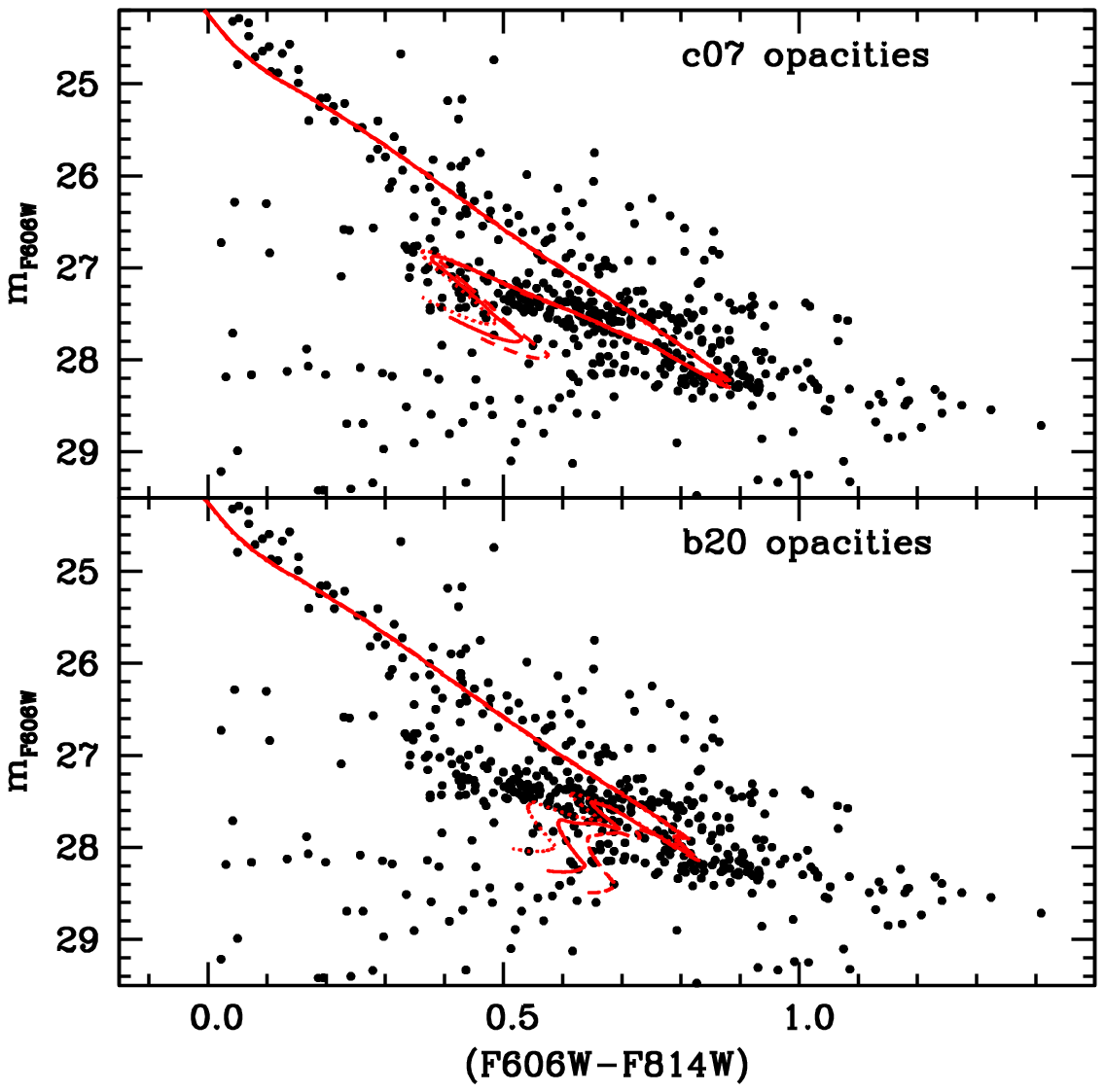}
      \caption{As the lower panel of Fig.~\ref{fig:nodist}, but  
      showing two sets of theoretical [Fe/H]=0.3 WD isochrones 
      with ${\rm ^{22}Ne}$ distillation calculated employing the c07 
      (upper panel) and b20 (lower panel) opacities, for ages equal to 8.0 (dotted lines), 8.5 (solid lines), and 9.0~Gyr (dashed lines), respectively.}
         \label{fig:isodist}
   \end{figure}

The models described in the previous section have been employed to calculate WD isochrones for both choices of opacity, and various ages, using the same 
progenitor metallicity, lifetimes, and initial-final mass relation described in Sect.~\ref{cluster}. Figure~\ref{fig:isodist} compares the CMD of isochrones 
with ages equal to 8, 8.5, and 9 Gyr, with the observed cooling sequence, 
after applying the same reddening and distance modulus as in  
Fig.~\ref{fig:nodist}.

There are a few striking features to notice in this comparison. First of all, 
the faint end of the 8.5~Gyr isochrones is shifted to brighter magnitudes compared to their counterparts in Fig.~\ref{fig:nodist}. This is the effect of the 
extra energy input due to the distillation of neon, that slows down the evolution of the models undergoing CO crystallization.

The second obvious feature is the change of shape of the faint part of the isochrones compared to the calculations without distillation, a consequence of the variation   
of $\Delta {\rm t_{cool}}$ induced by distillation with $M_{\rm WD}$.
The other important feature to notice is the weak effect of age (in the age range relevant to this cluster) on the CMD position and shape of the faint end of the isochrones, especially in the case of calculations with the c07 opacities. The reason is that at the luminosities corresponding to this isochrones' age range\footnote{Let's recall here that at any luminosity along a WD isochrone of age $t_{\rm iso}$ the following relation holds: $t_{\rm iso}=t_{\rm cool} + t_{\rm prog}$, where 
$t_{\rm cool}$ is the cooling time of the WD mass at that luminosity, and $t_{\rm prog}$ 
its progenitor lifetime.}, neon distillation is still efficient in the cores of 
all masses between $M_{\rm WD}\sim$0.6$M_{\odot}$ and $\sim$0.65-0.9$M_{\odot}$ (depending on the chosen opacities), and 
the extra energy gained in the process slows 
the cooling to a level that their luminosity is virtually unchanged. 

\begin{table}
\centering
\caption{\label{tab:delay} Cooling delay $\Delta t_{\rm cool}$ caused by ${\rm ^{22}Ne}$ 
distillation for different WD masses along the 8.5~Gyr isochrones  
of Fig.~\ref{fig:isodist}, calculated with the c07 and b20 opacities respectively.
}
{
\begin{tabular}{ccc}
\hline
$M_{WD} (M_{\odot})$ & $\Delta t_{\rm cool}^{\rm c07}$ (Gyr)  & $\Delta t_{\rm cool}^{\rm b20}$ (Gyr)\\  
\hline
0.61  & 1.20 & 2.00\\     
0.68  & 3.30 & 4.50\\       
0.77  & 4.50 & 4.20\\      
0.87  & 5.10 & 3.89\\       
1.00  & 2.43 & 2.52\\       
1.10  & 2.98 & 2.84\\     
  \hline
\end{tabular}
}
\end{table}

   \begin{figure}
     \centering
     \includegraphics[width=\columnwidth]{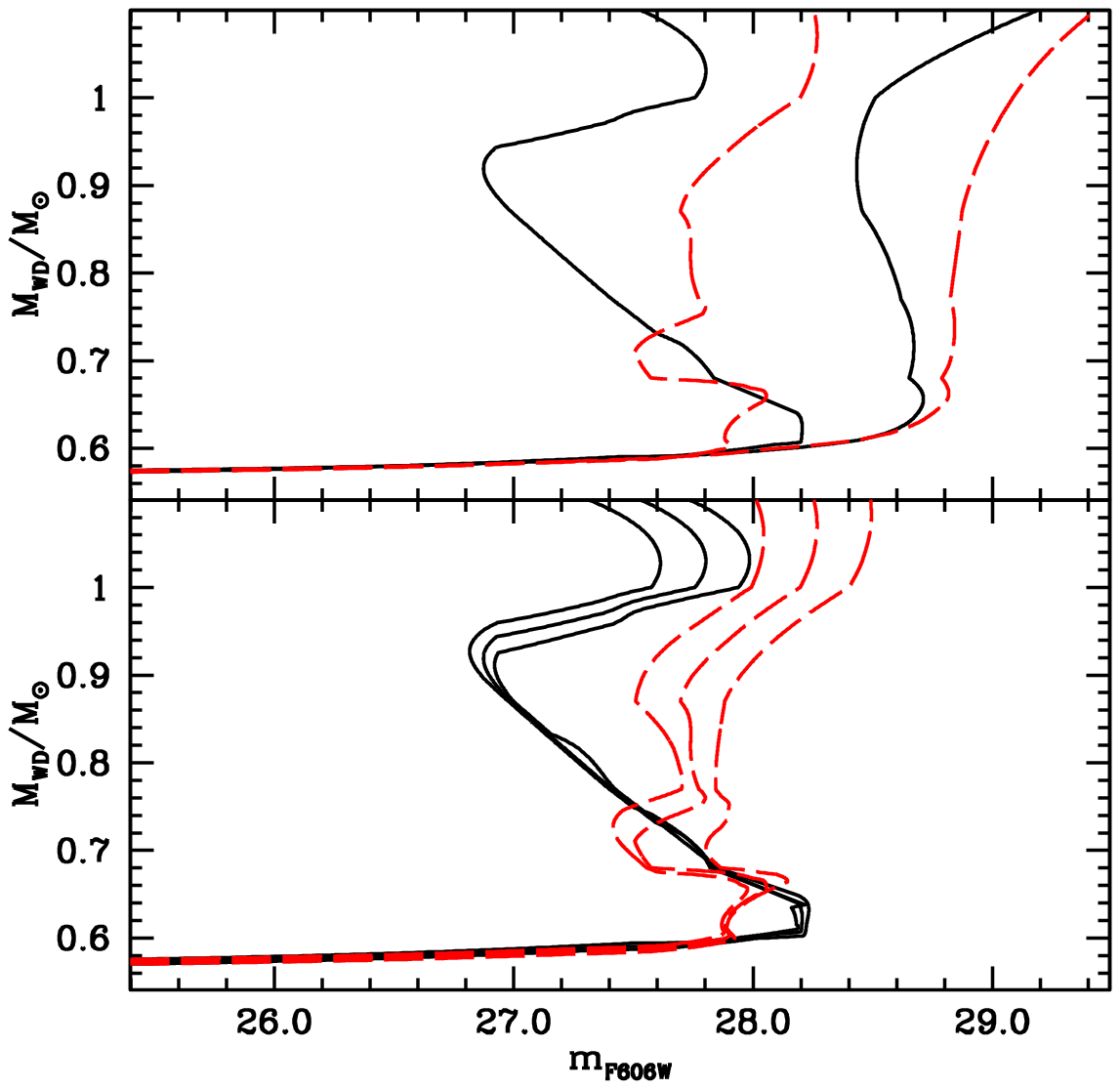}
      \caption{{\sl Upper panel:}
      Distribution of the evolving WD mass as a function of 
      the $m_{F606W}$ magnitude along the 8.5~Gyr isochrones with and without neon distillation. Black (solid) and red (long dashed) lines denote calculations with the c07 and b20 opacities, respectively. {\sl Lower panel:} Same as the upper panel, but for the 8.0, 8.5, and 9.0~Gyr isochrones of Fig.~\ref{fig:isodist} that all include 
      neon distillation, calculated with the c07 (solid lines) and b20 (dashed lines) opacities, respectively.}
         \label{fig:isomass}
   \end{figure}

All these points become even clearer when we consider  
Fig.~\ref{fig:isomass}, which displays the distribution of the evolving WD mass  
along the isochrones of Figs.~\ref{fig:nodist} and ~\ref{fig:isodist}. 
The upper panel shows the increase in the brightness of all masses at the faint end of the isochrones, caused by the cooling delays $\Delta t_{\rm cool}$ due to neon distillation, listed in Table~\ref{tab:delay} for representative masses. 
The shift in magnitude for a given $M_{\rm WD}$ depends on the value of $\Delta t_{\rm cool}$ --larger for 
higher $\Delta t_{\rm cool}$-- but also on the cooling speed when distillation is not included --larger shifts for higher cooling speed without distillation. 

   \begin{figure}
     \centering
     \includegraphics[width=\columnwidth]{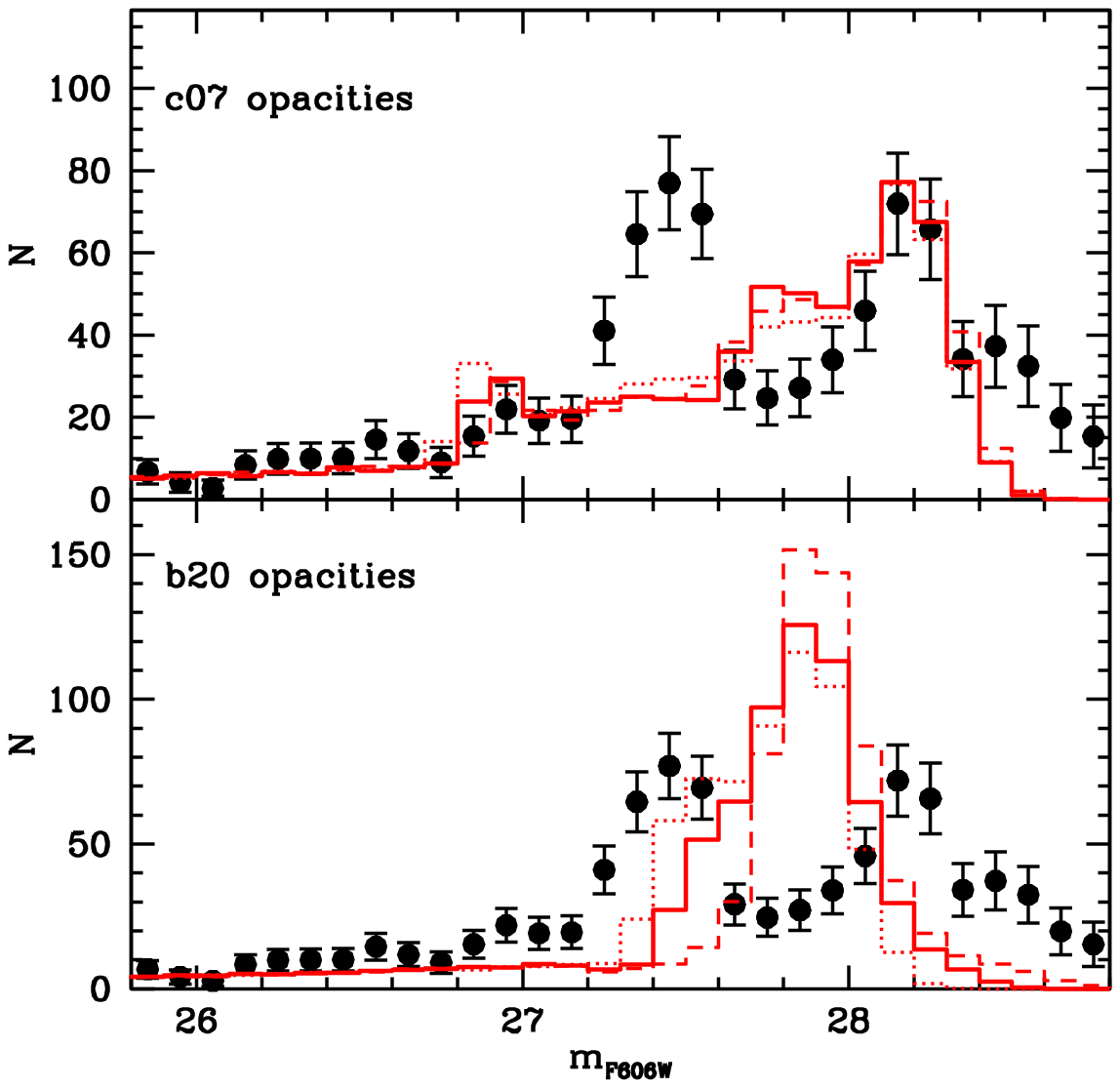}
      \caption{Comparison of the observed 
      WD LF of NGC~6791 (filled circles with error bars) with theoretical LFs calculated from the same isochrones of Fig.~\ref{fig:isodist}. Dotted, solid, and dashed lines correspond to ages equal to 8.0, 8.5, and 9.0~Gyr, respectively. The normalization of the theoretical LFs is arbitrary (but the total number of stars is the same in all the theoretical LFs).}
         \label{fig:lf}
   \end{figure}

The lower panel of Figure~\ref{fig:isomass} is similar to the upper panel, but it displays only isochrones which include neon distillation, with ages equal to 8.0, 8.5 and 9.0~Gyr, as in Fig.~\ref{fig:isodist}. It is shown very clearly that the brightness of all masses between $\sim$0.6$M_{\odot}$ and $\sim$0.65$M_{\odot}$ (for b20 opacities) or $\sim$0.9$M_{\odot}$ (for c07 opacities) is independent of the isochrone age, because these objects are still 
undergoing neon distillation in their cores, and their luminosity evolution essentially stalls.
Higher masses show a variation in brightness with changing isochrone age, because they are evolving past the completion of neon distillation, and the energy contribution of this process has vanished.

Are the isochrones including neon distillation able to reconcile the WD age with the main sequence age? Figure~\ref{fig:lf} shows more clearly than the CMD comparison that this is indeed the case, at 
least for models calculated with c07 opacities. We display here the 
observed LF of NGC~6791 WDs (bin size of 0.1~mag), together with the LFs calculated from our isochrones with neon distillation. We can see that in the case of calculations with the c07 
opacities the termination of the theoretical LF matches the observations.
The increase in the number of objects when approaching the bottom of the theoretical 
LF is caused by the increasing number of WDs when $M_{\rm WD}$ decreases from 
0.7-0.8$M_{\odot}$ to $\sim$0.6$M_{\odot}$ --the mass evolving at the faint end 
of the isochrone-- due to the initial-final mass relation and the power law with a negative exponent for the chosen progenitor mass function.
It is important to stress that in the case of the c07 isochrones the magnitude of the termination of the 
theoretical WD LF does not change if the progenitor mass function is changed --because the $m_{F606W}$ magnitude along the isochrone has 
an absolute maximum at $M_{\rm WD}\sim$0.60-0.64-- 
but the overall shape of the LF does.

   \begin{figure}
     \centering
     \includegraphics[width=\columnwidth]{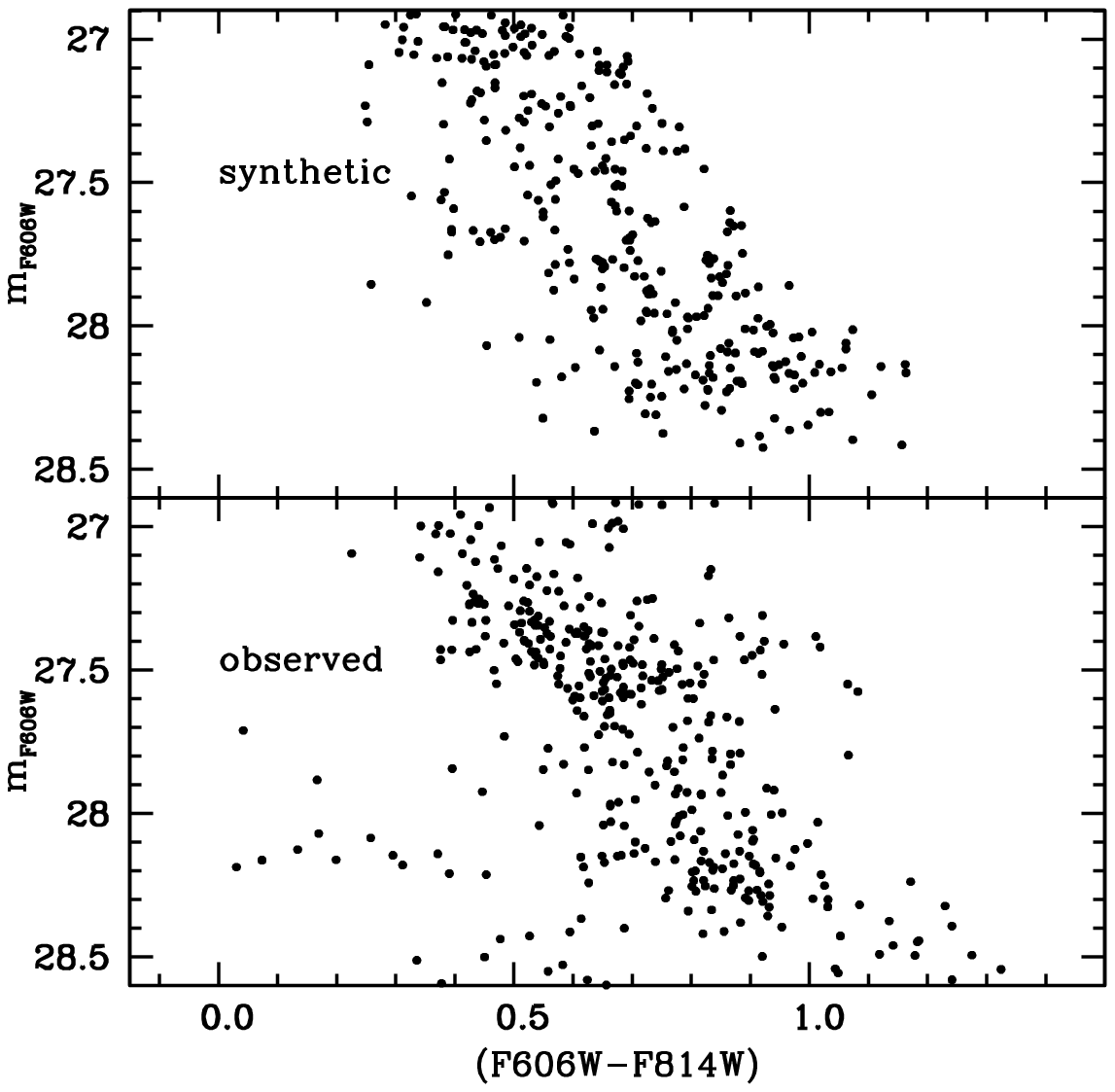}
      \caption{Comparison of the observed CMD of NGC~6791 WDs, with a synthetic CMD computed from the 8.5~Gyr 
      WD isochrone (c07 opacities) in the top panel of Fig.~\ref{fig:isodist} (see text for details).}
         \label{fig:hr1}
   \end{figure}

The different shape of the isochrones in Fig.~\ref{fig:isodist} compared to the case without neon distillation implies that their faint end covers a very small 
colour range, potentially in contrast with the observed broad cooling sequence at these magnitudes.
However, this small color range is not in disagreement with 
the observations once photometric errors are added. This is 
shown by Fig.~\ref{fig:hr1}, which displays a synthetic CMD for the 8.5~Gyr c07 isochrone, including the appropriate photometric errors. This synthetic CMD has been calculated as described in Sect.~\ref{cluster}, but we have then subtracted stars 
according to the completeness factors as a function of magnitude derived from the observations \citep{ngc6791}, to simulate the 
observed CMD also displayed in Fig.~\ref{fig:hr1} (the number of stars in both diagrams is the same).
The bottom end of the synthetic CMD also shows 
a wide colour range like the observations, entirely due to the large 1$\sigma$ error on the ($F606W-F814W$) that, e.g. at a reference $m_{F606W}$=28.0, amounts to $\sim$0.1~mag.

The comparison with the theoretical LFs from isochrones calculated with 
the b20 opacities is more complex, because of the different shape of the $M_{\rm WD}$-$m_{F606W}$ relation (and of the 
isochrones) compared to the previous c07 case.

Figure~\ref{fig:lf} shows that these 
LFs display a too-bright faint peak compared to the observations, by about 0.2~mag.
However, the situation is less clear-cut if we look 
again at Fig.~\ref{fig:isomass}. Due to the non-monotonic trend of the 
$m_{F606W}$ with increasing WD mass in the range between $\sim$0.6$M_{\odot}$ and $\sim$0.7$M_{\odot}$, along the b20 isochrones WD masses 
around 0.66$M_{\odot}$ have a magnitude $m_{F606W} \sim$28.1 at an age around 8.5-9.0~Gyr, consistent 
with the magnitude of the faint peak of the observed cluster WD LF. 
Due to the 
chosen progenitor mass function (and initial-final mass relation), in the actual LF the number of objects with this mass 
(coming from progenitors of $\sim$2.5$M_{\odot}$ according to the assumed 
initial-final mass relation) is lower than the number of objects 
with a mass around 0.6$M_{\odot}$ (coming from progenitors with mass equal to 
$\sim$1.5$M_{\odot}$) that populate the peak of the LF.

If the actual WD mass distribution in the cluster 
is instead, and somehow {\sl ad-hoc}, peaked around 0.66$M_{\odot}$ --due for example to  an initial-final mass relation and/or 
a progenitor mass function different from the one assumed 
in these calculations, and/or dynamical evolutionary effects that have affected 
the mass distribution of the objects in the observed cluster field-- 
the LF calculated with b20 opacities would show a peak consistent with 
the observed faint peak for ages between $\sim$8.5 and 9.0~Gyr.


\section{An observational test with {\sl JWST}}
\label{test}

Before closing the paper we discuss briefly an observational test that could confirm independently the efficiency of neon distillation in the interiors of this cluster's WDs, and potentially even help us identify which one is the more accurate treatment of electron degeneracy opacity, between the c07 and b20 results.

   \begin{figure}
     \centering
\includegraphics[width=\columnwidth]{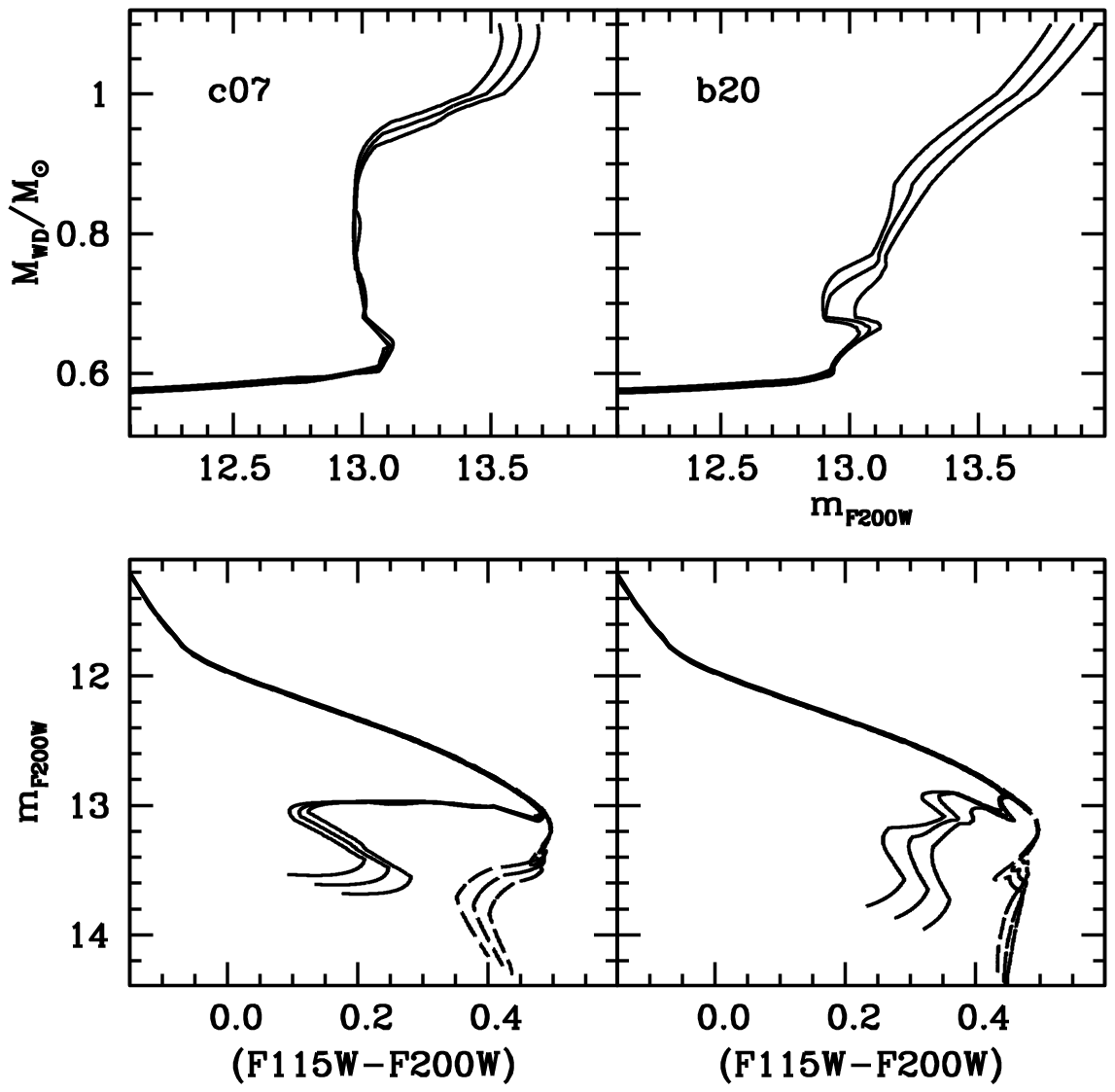}
      \caption{{\sl Left panels:} Distribution of the evolving WD mass as a function of the $JWST$ $F200W$ absolute magnitude along the 8.0, 8.5, and 9.0 Gyr [Fe/H]=0.3 isochrones with neon distillation, calculated with the c07 opacities (solid lines) and the corresponding isochrones in a $JWST$ CMD (solid lines in the lower panel) compared to the counterparts without neon distillation (dashed lines in the lower panel). {\sl Right panel:} As in the left panels, but for isochrones calculated with the b20 opacities.}
         \label{fig:jwstiso}
   \end{figure}

Figure~\ref{fig:jwstiso} displays the $JWST$/NIRCam ($F115W$,$F200W$) CMD of the same 
WD isochrones with and without distillation investigated so far, for 8.0. 8.5, and 9.0~Gyr, respectively.
Also in these filters, the difference in the shape between the two sets is remarkable.
When distillation is included, the isochrones calculated with the c07 opacities display --due to the behaviour of the bolometric corrections with $T_{\rm eff}$-- an extended 
(by 0.3-0.4~mag) horizontal sequence in the ($F115W-F200W$) colour, which is not 
present in the counterpart without distillation.
This horizontal sequence is populated by the objects with mass between $\sim$0.6$M_{\odot}$ and 0.9$M_{\odot}$, as shown also in Fig.~\ref{fig:jwstiso}. The same objects in the isochrone without distillation are distributed along a tilted sequence in this CMD, which is also much less extended in colour than the isochrone with distillation.
In the case of models calculated with the b20 opacities, the isochrones with 
distillation are more horizontal than those without, but less straight and with a narrower colour extension compared to the case of their c07 counterpart.

\begin{figure}
     \centering
\includegraphics[width=\columnwidth]{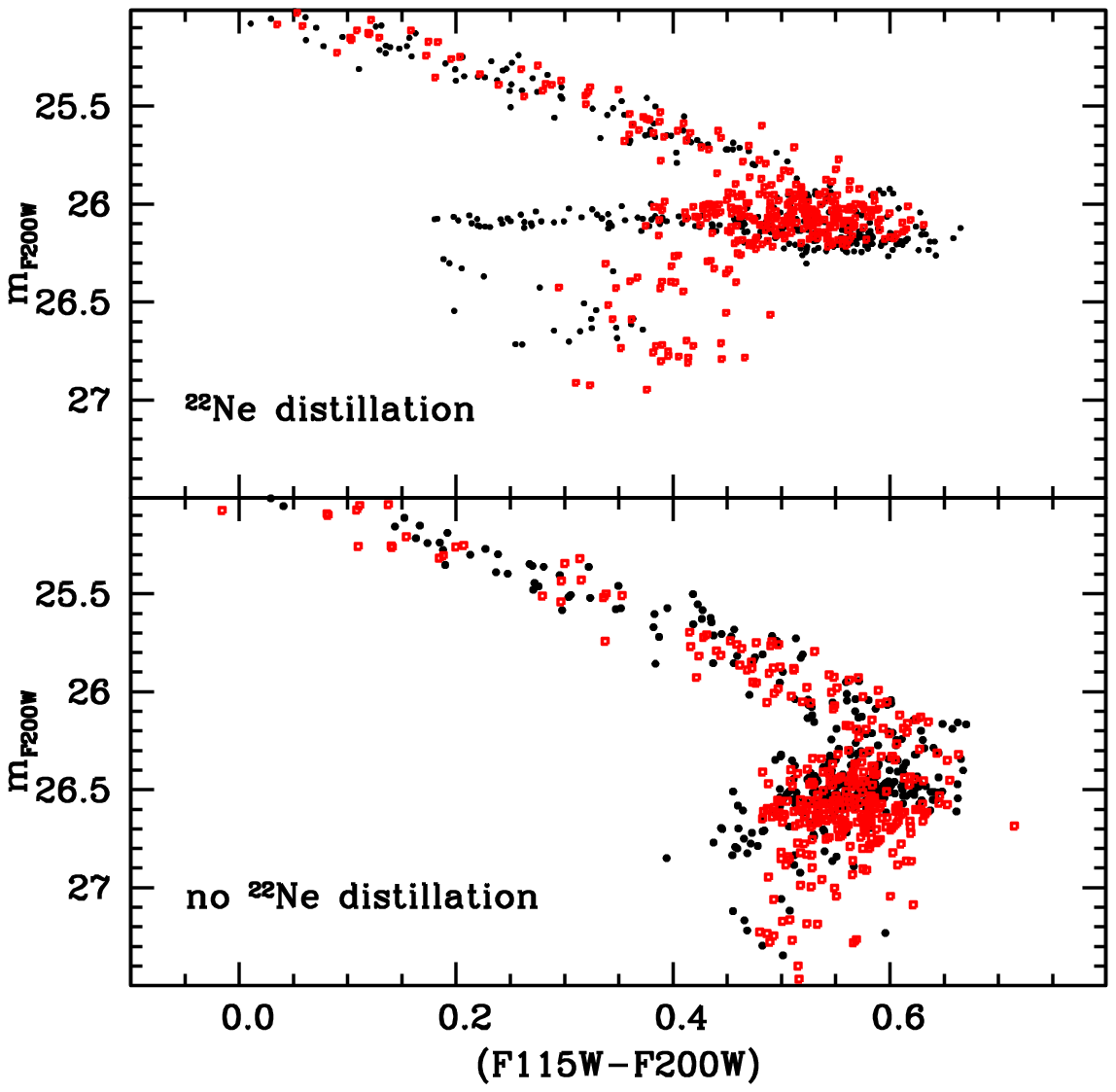}
      \caption{{\sl Upper panel:} Synthetic $JWST$ CMDs for 8.5~Gyr WD populations (see text for details) at the distance of NGC~6791, calculated from isochrones with neon distillation and either the c07 (black filled circles) or the b20 (red open squares) opacities.
       {\sl Lower panel:} As in the upper panel, but for calculations without 
       the inclusion of neon distillation.
       }
         \label{fig:jwstsynth}
   \end{figure}

Figure~\ref{fig:jwstsynth} displays the synthetic CMDs (350 stars each\footnote{The approximate number expected by taking into account the different field of view size and shape compared to $HST$/ACS, and the number of objects detected with the $HST$ observations}) obtained for the 8.5~Gyr isochrones with and without distillation, shifted to the apparent magnitudes and colours expected for NGC~6791 in these filters\footnote{We have used the extinction law by \citet{wang} for the $JWST$/NIRCam filters to transform the apparent distance modulus in $V$ and $E(B-V)$ adopted in our analysis, to the counterparts for our  
$JWST$ CMD.}, and 1$\sigma$ photometric errors in $F115W$ and $F200W$ equal to 0.03~mag.
For a given choice of the opacity the differences between the CMDs of the two populations (with and without distillation) are still quite obvious. It is also intriguing to notice 
that with these photometric errors, the synthetic populations calculated with distillation and the two different choices of opacity display 
a similar horizontal morphology at their faint well-populated branch, 
but with different colour extensions (the extension in the b20 CMD is about half of the c07 case, mirroring the differences seen 
between the corresponding isochrones in Fig.~\ref{fig:jwstiso}).

We conclude that an observed CMD with these photometric errors should be able to further confirm (or disprove) the efficiency of neon distillation in the interiors of NGC~6791 WDs, and also help us identify the more appropriate treatment of the electron conduction opacity in the regime at the transition from 
moderate to strong degeneracy.

To assess whether these observations are feasible, we employed the most updated online $JWST$ exposure time
calculator (ETC Version 3.0)\footnote{https://jwst.etc.stsci.edu/}, and 
estimated a required integration time of about 5\,hrs for each of the
two filters $F115W$ and $F200W$ to attain the signal-to-noise
ratio (SNR) of ~25-30 needed to achieve the random photometric error used in these synthetic CMDs.
For this calculation, we have assumed a black-body spectral energy distribution for a temperature of 5000~K (the approximate $T_{\rm eff}$ of the faintest WDs in the simulation), normalized at $m_{F115W}$=27.0 and $m_{F200W}$=26.7 for the expected end of the 
cooling sequence.
These numbers are also supported by our $JWST$ 
multi-filter calibrated photometry of the open cluster
NGC\,2506, available from the catalog publicly released by \citet{nard}.
In those observations, a total exposure time of just 43\,s in the filter $F115W$ allowed us to reach a SNR=$\sim$15-30 down to $m_{F115W}$=21.
According to an approximate calculation based just on the rescaling of the exposure times, we would expect to go $\sim$6 mags fainter in 3 hours with $F115W$, and similarly for the redder filter $F200W$.

\section{Summary and conclusions}
\label{discussion}

We have presented the first CO WD evolutionary models that self-consistently include the effect of neon distillation in addition to CO crystallization/phase separation and neon diffusion in the liquid phase. Our calculations have been performed including the c07 and b20 electron conduction opacities, respectively, and 
cover the full range of CO WD masses, for a progenitor metallicity twice-solar appropriate for the old open cluster NGC~6791.

We have shown that these state-of-the-art WD evolutionary models produce LFs that can match the faint end of the observed WD LF in NGC~6791, for ages consistent with the narrow range determined from eclipsing binary stars and the main sequence turn-off. Without the inclusion of distillation the termination of 
the theoretical WD cooling sequences is too faint compared to the observations.

All cluster WDs with masses between $\sim$0.6$M_{\odot}$ and $\sim$0.65-0.9$M_{\odot}$ (for the b20 and c07 opacities, respectively) are undergoing neon distillation and the shape of the isochrones is very different from the case without distillation. Also remarkable is the lack of evolution in the 
brightness of the LF faint end when distillation is included, which is due to 
the luminosity evolution of the models stalling whilst distillation is ongoing.

This means that at the metallicity and age of NGC~6791, WD age-dating on its own may not be very precise when neon distillation is included in the models' computation, particularly for models calculated with the c07 opacities. 
In this case, we find that theoretical LFs with ages ranging from 7 to 10~Gyr that include the photometric 
errors of the cluster data, all display the same magnitude of the faint 
peak at the termination of the cooling sequence.
If we did not have tight constraints on NGC~6791 age from the main sequence, the WD LF --with neon distillation included-- would be consistent with an age range of 4~Gyr. Even after reducing the photometric errors to just 0.02-0.03 mag 
across the whole LF, increasing the number of objects, and reducing the bin size by half, the magnitude of the termination of the theoretical LF is still basically the same between 7 and 10 Gyr. 
This is at odds with the case of models without neon distillation, which 
produced LFs with varying peak magnitudes across this age range.

In the case of models calculated with the b20 opacities this age degeneracy is much mitigated, because only the LFs in the age range 
between 8.5 and 9 Gyr display the same magnitude of the peak at the termination of the cooling sequence. However, in this case, we need a fine-tuned mass distribution (peaked around $M_{\rm WD}\sim$0.66) to match the observed faint peak of the cluster WD LF.

We have also proposed and described new {\sl JWST} observations that can independently prove the efficiency of neon distillation in the interiors of NGC~6791 WDs, 
based on the shape of the cluster cooling sequence in appropriately chosen {\sl JWST} CMDs.
An important byproduct of these proposed observations is their potential also to help solve the current uncertainties on the electron conduction opacities for the modelling of WD envelopes.

Finally, as we mentioned in the Introduction, all other metals in the CO WD cores are at least one order of magnitude less abundant (in mass fraction) than $^{22}{\rm Ne}$, and their contribution to the energy budget --hence cooling times-- might be expected to be small, if not negligible. Indeed this seems to be the case when we consider $^{56}{\rm Fe}$, the most abundant element after $^{22}{\rm Ne}$, with a 
mass fraction equal to only about 9\% of the $^{22}{\rm Ne}$ abundance \citep[see, e.g.][]{bastiiacwd}.  
\citet{bastiiacwd} have made use of the results 
by \citet{caplanFe} about the phase diagram of a COFe mixture, to show that the sedimentation of Fe in the CO core during phase separation increases the cooling times of CO WD models at the level of at most 100-200~Myr for a metallicity $Z$=0.04 --even higher than that of NGC~6791-- a negligible contribution compared to the effect of neon diffusion and distillation.

Our modeling work has highlighted uncertainties related to the stopping/starting condition for distillation that should be investigated in future work. In particular, it is unclear whether $^{22}$Ne-rich shells can be formed after distillation is completed in the inner core. Future work should also try to improve upon the current treatment of composition change during distillation, which is currently assumed to be linear with respect to the Coulomb coupling parameter.

\begin{acknowledgements}
We thank our referee for comments and suggestions that have improved the presentation of our results.
MS acknowledges support from The Science and Technology Facilities
Council Consolidated Grant ST/V00087X/1. SB thanks the Banting Postdoctoral Fellowship and CITA National Fellowship programs for support and is grateful to Matt Caplan for useful discussions. SC acknowledges financial support from PRIN-MIUR-22: CHRONOS: adjusting the clock(s) to unveil the CHRONO-chemo-dynamical Structure of the Galaxy” (PI: S. Cassisi) finanziato dall'Unione Europea - Next Generation EU, and
from INFN (Iniziativa specifica TAsP). 
\end{acknowledgements}

\bibliographystyle{aa}
\bibliography{WDpaper}

\begin{thebibliography}{49}
\expandafter\ifx\csname natexlab\endcsname\relax\def\natexlab#1{#1}\fi

\bibitem[{{Althaus} {et~al.}(2010){Althaus}, {Garc{\'\i}a-Berro}, {Renedo}, {Isern}, {C{\'o}rsico}, \& {Rohrmann}}]{a10}
{Althaus}, L.~G., {Garc{\'\i}a-Berro}, E., {Renedo}, I., {et~al.} 2010, \apj, 719, 612

\bibitem[{{Bauer}(2023)}]{bauer}
{Bauer}, E.~B. 2023, \apj, 950, 115

\bibitem[{{Bedin} {et~al.}(2005){Bedin}, {Cassisi}, {Castelli}, {Piotto}, {Anderson}, {Salaris}, {Momany}, \& {Pietrinferni}}]{bedin05}
{Bedin}, L.~R., {Cassisi}, S., {Castelli}, F., {et~al.} 2005, \mnras, 357, 1038

\bibitem[{{Bedin} {et~al.}(2008{\natexlab{a}}){Bedin}, {King}, {Anderson}, {Piotto}, {Salaris}, {Cassisi}, \& {Serenelli}}]{ngc6791}
{Bedin}, L.~R., {King}, I.~R., {Anderson}, J., {et~al.} 2008{\natexlab{a}}, \apj, 678, 1279

\bibitem[{{Bedin} {et~al.}(2015){Bedin}, {Salaris}, {Anderson}, {Cassisi}, {Milone}, {Piotto}, {King}, \& {Bergeron}}]{ngc6819}
{Bedin}, L.~R., {Salaris}, M., {Anderson}, J., {et~al.} 2015, \mnras, 448, 1779

\bibitem[{{Bedin} {et~al.}(2019){Bedin}, {Salaris}, {Anderson}, {Libralato}, {Apai}, {Nardiello}, {Rich}, {Bellini}, {Dieball}, {Bergeron}, {Burgasser}, {Milone}, \& {Marino}}]{ngc6752}
{Bedin}, L.~R., {Salaris}, M., {Anderson}, J., {et~al.} 2019, \mnras, 488, 3857

\bibitem[{{Bedin} {et~al.}(2010){Bedin}, {Salaris}, {King}, {Piotto}, {Anderson}, \& {Cassisi}}]{ngc2158}
{Bedin}, L.~R., {Salaris}, M., {King}, I.~R., {et~al.} 2010, \apjl, 708, L32

\bibitem[{{Bedin} {et~al.}(2009){Bedin}, {Salaris}, {Piotto}, {Anderson}, {King}, \& {Cassisi}}]{m4}
{Bedin}, L.~R., {Salaris}, M., {Piotto}, G., {et~al.} 2009, \apj, 697, 965

\bibitem[{{Bedin} {et~al.}(2008{\natexlab{b}}){Bedin}, {Salaris}, {Piotto}, {Cassisi}, {Milone}, {Anderson}, \& {King}}]{ngc6791bin}
{Bedin}, L.~R., {Salaris}, M., {Piotto}, G., {et~al.} 2008{\natexlab{b}}, \apjl, 679, L29

\bibitem[{{Bellini} {et~al.}(2010){Bellini}, {Bedin}, {Piotto}, {Salaris}, {Anderson}, {Brocato}, {Ragazzoni}, {Ortolani}, {Bonanos}, {Platais}, {Gilliland}, {Raimondo}, {Bragaglia}, {Tosi}, {Gallozzi}, {Testa}, {Kochanek}, {Giallongo}, {Baruffolo}, {Farinato}, {Diolaiti}, {Speziali}, {Carraro}, \& {Yadav}}]{m67}
{Bellini}, A., {Bedin}, L.~R., {Piotto}, G., {et~al.} 2010, \aap, 513, A50

\bibitem[{{Bildsten} \& {Hall}(2001)}]{bh01}
{Bildsten}, L. \& {Hall}, D.~M. 2001, \apjl, 549, L219

\bibitem[{{Blaes} {et~al.}(1990){Blaes}, {Blandford}, {Madau}, \& {Koonin}}]{blaes}
{Blaes}, O., {Blandford}, R., {Madau}, P., \& {Koonin}, S. 1990, \apj, 363, 612

\bibitem[{{Blouin} \& {Daligault}(2021)}]{bd21}
{Blouin}, S. \& {Daligault}, J. 2021, \pre, 103, 043204

\bibitem[{{Blouin} {et~al.}(2021){Blouin}, {Daligault}, \& {Saumon}}]{b21}
{Blouin}, S., {Daligault}, J., \& {Saumon}, D. 2021, \apjl, 911, L5

\bibitem[{{Blouin} {et~al.}(2020){Blouin}, {Shaffer}, {Saumon}, \& {Starrett}}]{b20}
{Blouin}, S., {Shaffer}, N.~R., {Saumon}, D., \& {Starrett}, C.~E. 2020, \apj, 899, 46

\bibitem[{{Brogaard} {et~al.}(2011){Brogaard}, {Bruntt}, {Grundahl}, {Clausen}, {Frandsen}, {Vandenberg}, \& {Bedin}}]{brogaard11}
{Brogaard}, K., {Bruntt}, H., {Grundahl}, F., {et~al.} 2011, \aap, 525, A2

\bibitem[{{Brogaard} {et~al.}(2021){Brogaard}, {Grundahl}, {Sandquist}, {Slumstrup}, {Jensen}, {Thomsen}, {J{\o}rgensen}, {Larsen}, {Bj{\o}rn}, {S{\o}rensen}, {Bruntt}, {Arentoft}, {Frandsen}, {Jessen-Hansen}, {Orosz}, {Mathieu}, {Geller}, {Ryde}, {Stello}, {Meibom}, \& {Platais}}]{brogaard}
{Brogaard}, K., {Grundahl}, F., {Sandquist}, E.~L., {et~al.} 2021, \aap, 649, A178

\bibitem[{{Brogaard} {et~al.}(2012){Brogaard}, {VandenBerg}, {Bruntt}, {Grundahl}, {Frandsen}, {Bedin}, {Milone}, {Dotter}, {Feiden}, {Stetson}, {Sandquist}, {Miglio}, {Stello}, \& {Jessen-Hansen}}]{brogaard12}
{Brogaard}, K., {VandenBerg}, D.~A., {Bruntt}, H., {et~al.} 2012, \aap, 543, A106

\bibitem[{{Camisassa} {et~al.}(2016){Camisassa}, {Althaus}, {C{\'o}rsico}, {Vinyoles}, {Serenelli}, {Isern}, {Miller Bertolami}, \& {Garc{\'\i}a{\textendash}Berro}}]{camisassa16}
{Camisassa}, M.~E., {Althaus}, L.~G., {C{\'o}rsico}, A.~H., {et~al.} 2016, \apj, 823, 158

\bibitem[{{Caplan} {et~al.}(2021){Caplan}, {Freeman}, {Horowitz}, {Cumming}, \& {Bellinger}}]{caplanFe}
{Caplan}, M.~E., {Freeman}, I.~F., {Horowitz}, C.~J., {Cumming}, A., \& {Bellinger}, E.~P. 2021, arXiv e-prints, arXiv:2108.11389

\bibitem[{{Caplan} {et~al.}(2020){Caplan}, {Horowitz}, \& {Cumming}}]{caplan20}
{Caplan}, M.~E., {Horowitz}, C.~J., \& {Cumming}, A. 2020, \apjl, 902, L44

\bibitem[{{Cassisi} {et~al.}(2007){Cassisi}, {Potekhin}, {Pietrinferni}, {Catelan}, \& {Salaris}}]{cas07}
{Cassisi}, S., {Potekhin}, A.~Y., {Pietrinferni}, A., {Catelan}, M., \& {Salaris}, M. 2007, \apj, 661, 1094

\bibitem[{{Cassisi} {et~al.}(2021){Cassisi}, {Potekhin}, {Salaris}, \& {Pietrinferni}}]{cpsp}
{Cassisi}, S., {Potekhin}, A.~Y., {Salaris}, M., \& {Pietrinferni}, A. 2021, \aap, 654, A149

\bibitem[{{Cukanovaite} {et~al.}(2023){Cukanovaite}, {Tremblay}, {Toonen}, {Temmink}, {Manser}, {O'Brien}, \& {McCleery}}]{cuk}
{Cukanovaite}, E., {Tremblay}, P.~E., {Toonen}, S., {et~al.} 2023, \mnras, 522, 1643

\bibitem[{{Cummings} {et~al.}(2018){Cummings}, {Kalirai}, {Tremblay}, {Ramirez-Ruiz}, \& {Choi}}]{cummings}
{Cummings}, J.~D., {Kalirai}, J.~S., {Tremblay}, P.~E., {Ramirez-Ruiz}, E., \& {Choi}, J. 2018, \apj, 866, 21

\bibitem[{{Deloye} \& {Bildsten}(2002)}]{db02}
{Deloye}, C.~J. \& {Bildsten}, L. 2002, \apj, 580, 1077

\bibitem[{{Garc{\'\i}a-Berro} {et~al.}(2008){Garc{\'\i}a-Berro}, {Althaus}, {C{\'o}rsico}, \& {Isern}}]{gb08}
{Garc{\'\i}a-Berro}, E., {Althaus}, L.~G., {C{\'o}rsico}, A.~H., \& {Isern}, J. 2008, \apj, 677, 473

\bibitem[{{Garc{\'{\i}}a-Berro} {et~al.}(2010){Garc{\'{\i}}a-Berro}, {Torres}, {Althaus}, {Renedo}, {Lor{\'e}n-Aguilar}, {C{\'o}rsico}, {Rohrmann}, {Salaris}, \& {Isern}}]{garciaberro}
{Garc{\'{\i}}a-Berro}, E., {Torres}, S., {Althaus}, L.~G., {et~al.} 2010, \nat, 465, 194

\bibitem[{{Goldsbury} {et~al.}(2012){Goldsbury}, {Heyl}, {Richer}, {Bergeron}, {Dotter}, {Kalirai}, {MacDonald}, {Rich}, {Stetson}, {Tremblay}, \& {Woodley}}]{ratecooling}
{Goldsbury}, R., {Heyl}, J., {Richer}, H.~B., {et~al.} 2012, \apj, 760, 78

\bibitem[{{Hansen}(2005)}]{hansen05}
{Hansen}, B. M.~S. 2005, \apj, 635, 522

\bibitem[{{Hansen} {et~al.}(2007){Hansen}, {Anderson}, {Brewer}, {Dotter}, {Fahlman}, {Hurley}, {Kalirai}, {King}, {Reitzel}, {Richer}, {Rich}, {Shara}, \& {Stetson}}]{ngc6397}
{Hansen}, B. M.~S., {Anderson}, J., {Brewer}, J., {et~al.} 2007, \apj, 671, 380

\bibitem[{{Hansen} {et~al.}(2004){Hansen}, {Richer}, {Fahlman}, {Stetson}, {Brewer}, {Currie}, {Gibson}, {Ibata}, {Rich}, \& {Shara}}]{m4hansen}
{Hansen}, B. M.~S., {Richer}, H.~B., {Fahlman}, G.~G., {et~al.} 2004, \apjs, 155, 551

\bibitem[{{Hidalgo} {et~al.}(2018){Hidalgo}, {Pietrinferni}, {Cassisi}, {Salaris}, {Mucciarelli}, {Savino}, {Aparicio}, {Silva Aguirre}, \& {Verma}}]{bastiiacss}
{Hidalgo}, S.~L., {Pietrinferni}, A., {Cassisi}, S., {et~al.} 2018, \apj, 856, 125

\bibitem[{{Isern} {et~al.}(1991){Isern}, {Hernanz}, {Mochkovitch}, \& {Garcia-Berro}}]{isern1991}
{Isern}, J., {Hernanz}, M., {Mochkovitch}, R., \& {Garcia-Berro}, E. 1991, \aap, 241, L29

\bibitem[{{Isern} {et~al.}(2022){Isern}, {Torres}, \& {Rebassa-Mansergas}}]{isernphys}
{Isern}, J., {Torres}, S., \& {Rebassa-Mansergas}, A. 2022, Frontiers in Astronomy and Space Sciences, 9, 6

\bibitem[{{Kilic} {et~al.}(2017){Kilic}, {Munn}, {Harris}, {von Hippel}, {Liebert}, {Williams}, {Jeffery}, \& {DeGennaro}}]{kilic17}
{Kilic}, M., {Munn}, J.~A., {Harris}, H.~C., {et~al.} 2017, \apj, 837, 162

\bibitem[{{Nardiello} {et~al.}(2023){Nardiello}, {Bedin}, {Griggio}, {Salaris}, {Scalco}, \& {Cassisi}}]{nard}
{Nardiello}, D., {Bedin}, L.~R., {Griggio}, M., {et~al.} 2023, \mnras, 525, 2585

\bibitem[{{Oswalt} {et~al.}(1996){Oswalt}, {Smith}, {Wood}, \& {Hintzen}}]{oswalt}
{Oswalt}, T.~D., {Smith}, J.~A., {Wood}, M.~A., \& {Hintzen}, P. 1996, \nat, 382, 692

\bibitem[{{Richer} {et~al.}(1998){Richer}, {Fahlman}, {Rosvick}, \& {Ibata}}]{m67richer}
{Richer}, H.~B., {Fahlman}, G.~G., {Rosvick}, J., \& {Ibata}, R. 1998, \apjl, 504, L91

\bibitem[{{Salaris} {et~al.}(2022){Salaris}, {Cassisi}, {Pietrinferni}, \& {Hidalgo}}]{bastiiacwd}
{Salaris}, M., {Cassisi}, S., {Pietrinferni}, A., \& {Hidalgo}, S. 2022, \mnras, 509, 5197

\bibitem[{{Saumon} {et~al.}(2022){Saumon}, {Blouin}, \& {Tremblay}}]{saumon22}
{Saumon}, D., {Blouin}, S., \& {Tremblay}, P.-E. 2022, \physrep, 988, 1

\bibitem[{{Segretain}(1996)}]{segretain1996}
{Segretain}, L. 1996, \aap, 310, 485

\bibitem[{{Segretain} {et~al.}(1994){Segretain}, {Chabrier}, {Hernanz}, {Garcia-Berro}, {Isern}, \& {Mochkovitch}}]{segretain94}
{Segretain}, L., {Chabrier}, G., {Hernanz}, M., {et~al.} 1994, \apj, 434, 641

\bibitem[{{Tononi} {et~al.}(2019){Tononi}, {Torres}, {Garc{\'\i}a-Berro}, {Camisassa}, {Althaus}, \& {Rebassa-Mansergas}}]{tononi}
{Tononi}, J., {Torres}, S., {Garc{\'\i}a-Berro}, E., {et~al.} 2019, \aap, 628, A52

\bibitem[{{Torres} \& {Garc{\'\i}a-Berro}(2016)}]{torres16}
{Torres}, S. \& {Garc{\'\i}a-Berro}, E. 2016, \aap, 588, A35

\bibitem[{{von Hippel}(2005)}]{vonHippel}
{von Hippel}, T. 2005, \apj, 622, 565

\bibitem[{{Wang} \& {Chen}(2019)}]{wang}
{Wang}, S. \& {Chen}, X. 2019, \apj, 877, 116

\bibitem[{{Winget} {et~al.}(1987){Winget}, {Hansen}, {Liebert}, {van Horn}, {Fontaine}, {Nather}, {Kepler}, \& {Lamb}}]{winget}
{Winget}, D.~E., {Hansen}, C.~J., {Liebert}, J., {et~al.} 1987, \apjl, 315, L77

\bibitem[{{Winget} {et~al.}(2009){Winget}, {Kepler}, {Campos}, {Montgomery}, {Girardi}, {Bergeron}, \& {Williams}}]{winget3}
{Winget}, D.~E., {Kepler}, S.~O., {Campos}, F., {et~al.} 2009, \apjl, 693, L6

\end{thebibliography}

\end{document}